\documentclass[aps,pra,groupedaddress]{revtex4-1}

\usepackage{graphicx}   
\usepackage{dcolumn}   
\usepackage{bm}   
\usepackage{epstopdf}
\usepackage{hyperref}   

\begin{document}

\preprint{}

\title{Comparison of ancilla preparation and measurement procedures for the
Steane [[7,1,3]] code on a model ion trap quantum computer}

\author{Yu Tomita} \author{Mauricio Guti\'{e}rrez} \author{Chingiz Kabytayev} \author{Kenneth R.
Brown} \altaffiliation{Author to whom correspondence should be addressed.
Electronic mail:kenbrown@gatech.edu } \affiliation{Schools of Chemistry and
Biochemistry; Computational Science and Engineering; and Physics, Georgia
Institute of Technology, Atlanta, Georgia 30332, USA}

\author{M. R. Hutsel} \author{A. P. Morris} \author{Kelly E. Stevens} \author{G. Mohler} \affiliation{Georgia
Tech Research Institute, Atlanta, Georgia 30332, USA}

\date{\today}

\begin{abstract}

We schedule the Steane [[7,1,3]] error correction on a model ion trap architecture
with ballistic transport. We compare the level one error rates for syndrome
extraction using the Shor method of ancilla prepared in verified cat states to the DiVincenzo-Aliferis method without verification. The study examines how the quantum error correction circuit latency and error vary with
the number of available ancilla and the choice of protocol for ancilla
preparation and measurement. We find that with few exceptions the DiVincenzo-Aliferis method without cat state verification outperforms the standard Shor method.  We also find that additional ancilla always reduces the latency but does not significantly change
the error due to the high memory fidelity.

\end{abstract}


\maketitle

\section{\label{sec:Int}Introduction}
The reliability of a fault-tolerant quantum computation depends on not only the
choice of error correction code but also the methods used for syndrome
extraction, state preparation, and error decoding. These choices can be
compared at an abstract level of quantum circuits and depolarizing channels,
but realistic quantum information devices will have error rates that depend on
circuit elements as well as limited connectivity for applying two-qubit gates
\cite{svore2007}. Topological codes have an advantage in that they are naturally suited
to nearest-neighbor architectures \cite{RaussendorfPRL2007,FowlerPRA2009}. Concatenated code error correction
procedures require additional resources to map these circuits onto local
architectures which leads to a reduced error threshold relative to the abstract
model \cite{crossmsthesis2005, svore2007}. Still, these codes offer potential
benefits over topological codes for systems with low-error rates and fast
communication between distant qubits by ballistic transport or interaction with
flying qubits.

 The extraction of syndromes requires the preparation and measurement of fresh
ancilla states.  This process is what allows us to remove the entropy from the
quantum system \cite{Aharonov1996}.  One question that arises is how many extra
qubits should one dedicate for ancilla.  Consider the Steane [[7,1,3]] code~\cite{Steane1996} using
the Shor method for syndrome extraction~\cite{DivincenzoPRL1996} based on verified cat states.  Each cat
state contains four qubits; six syndrome measurements are required,
suggesting that between 4 and 24 ancilla qubits could be used.  The proper
balance of ancilla resources depends on the device details and the error rates of the
physical operations. For most quantum information devices, measurement is the
slowest operation. It has been shown that in a nonequiprobable error environment where
Z type error is dominant, the fidelity of the Shor state may decreas with verification~\cite{Weinstein2011, Weinstein2012}.
To avoid bottlenecks due to the measurements used to verify
cat states, DiVincenzo and Aliferis \cite{divincenzo2007} proposed a method that
does not require verification of ancilla states. Here we compare these methods 
on a model ion trap quantum computer.

The ion trap architecture is a promising basis for quantum computation and have already
demonstrated long coherence times and high fidelity operations.  A scalable
architecture has been proposed based on shuttling ions between traps \cite{KMW} and work
is ongoing to implement this architecture experimentally \cite{SchulzNJP2008,SplattNJP2009,AminiNJP2010,HughesCP2011,MoehringNJP2011, BlakestadPRA2011,TrueNJP2011,DoretNJP2012,wright2012}. This framework has
been the basis for a number of studies on the resource requirements for
implementing large quantum algorithms \cite{kubi2008, MetodiMicro, ClarkPRA2009} and has also been considered as the
elementary logical unit of hybrid schemes using photonic interconnects \cite{musiqc}.

While an arbitrarily well-connected ion trap layout can be envisioned, such that
there is little fear of collision or
backlogs, this is not realistic given current technology.  The ion trap layout,
for example, is a grid of narrow paths where no ion may pass by another.  Performing
multiple two-qubit gates efficiently becomes problematic.  There will be a
limited number of interaction zones, and the paths to reach them will be
obstructed by other qubits which adds non-trivial transport time in addition to 
the time required to execute gates.

This introduces the issue of latency which is defined here as the total amount
of time experienced by qubits after physical state preparation.  Latency
includes qubit transport times, gate times, and idle times due to traffic in
the layout.  When mapping a quantum circuit to a series of device operations for a layout
with limited connectivity, resources dedicated to the transport of qubit
information quickly come to dominate the cost of algorithm execution~\cite{kubi2008}. The goal then becomes to find a schedule of qubit operations
that reduces latency as much as possible, both to make operation times feasible
for large algorithms and to reduce memory errors due to ever-present
environmental noise.  Parallelization of operations is one of the most direct
ways to reduce latency and is the focus of this work.

One simple way to increase the
parallelizability is to prepare additional ancillary qubits ahead of time in states
needed by the computation.  Just as in classical computing, this is a trade-off
between memory and latency.  Maximum parallelization may call for the
simultaneous creation and preparation of multiple ancilla sets (low latency),
but this results in ``stale'' ancilla that may suffer logical errors before
being used (poor memory performance).   Both of these factors can be calculated
quantitatively.  Total latency can be calculated given a layout, a schedule of
gates, and a set of operation times (gate time, qubit speed, measurement time,
etc.).  Logical error can be characterized in terms of fidelity, or alternatively in
terms of the qubit error rate.  In general, it will increase with increasing latency.  Using
these calculations, we can study the effect of additional resources
on the error rate of the overall algorithm execution.

The impact of ancilla preparation on overhead has been previously studied for
both individual logical qubits~\cite{salas2004, nada2013} and large-scale quantum
computation~\cite{kubi2008}.  The individual logical qubit studies done for
the Steane [[7,1,3]] code assumed an abstracted layout. Although the studies did consider memory errors due to gate operation times, they did not include the additional errors due to movement latency.    The large-scale
study looked at ion trap layouts holding large numbers of logical qubits, and
found that ancilla generation was the primary performance bottleneck.  The
bottleneck was removed by creating regions dedicated to ancilla preparation and
recycling.  Our approach flows in part from these prior studies; here, multiple
ancilla blocks are assigned to individual logical qubits, and two different
ancilla encodings are employed.  Once an ancilla block size and encoding are
chosen, execution of the Steane code is simulated using a software design tool.
The design tool is used to include realistic latency and scheduling
bottlenecks, pointing towards the most practical ancilla encoding and block
size. Our study focuses on a single round of Steane-code quantum error
correction on a model ion trap architecture as a function of ancilla
encoding/decoding and ancilla resources.

\section{\label{sec:Meth}Methods}

\subsection{Cat state syndrome extraction and the Steane [[7,1,3]] code} The
Steane [[7,1,3]] code is the best-known of the Calderbank-Shor-Steane codes
\cite{CSS}.  It encodes one logical qubit into seven physical qubits.  The resulting
logical states $|0\rangle$ and $|1\rangle$ have a Hamming distance of three and the code is able
to detect and correct up to one physical bit-flip error and one physical phase-flip error.  The Steane code has
been widely studied and has been shown to have a threshold in the range between $10^{-3}$ and
$10^{-6}$ \cite{svore2006, metodi2005, Steane2003} which makes it suitable for
fault-tolerant quantum error correction (FTQEC). 

\begin{figure}
\includegraphics[scale=1.0]{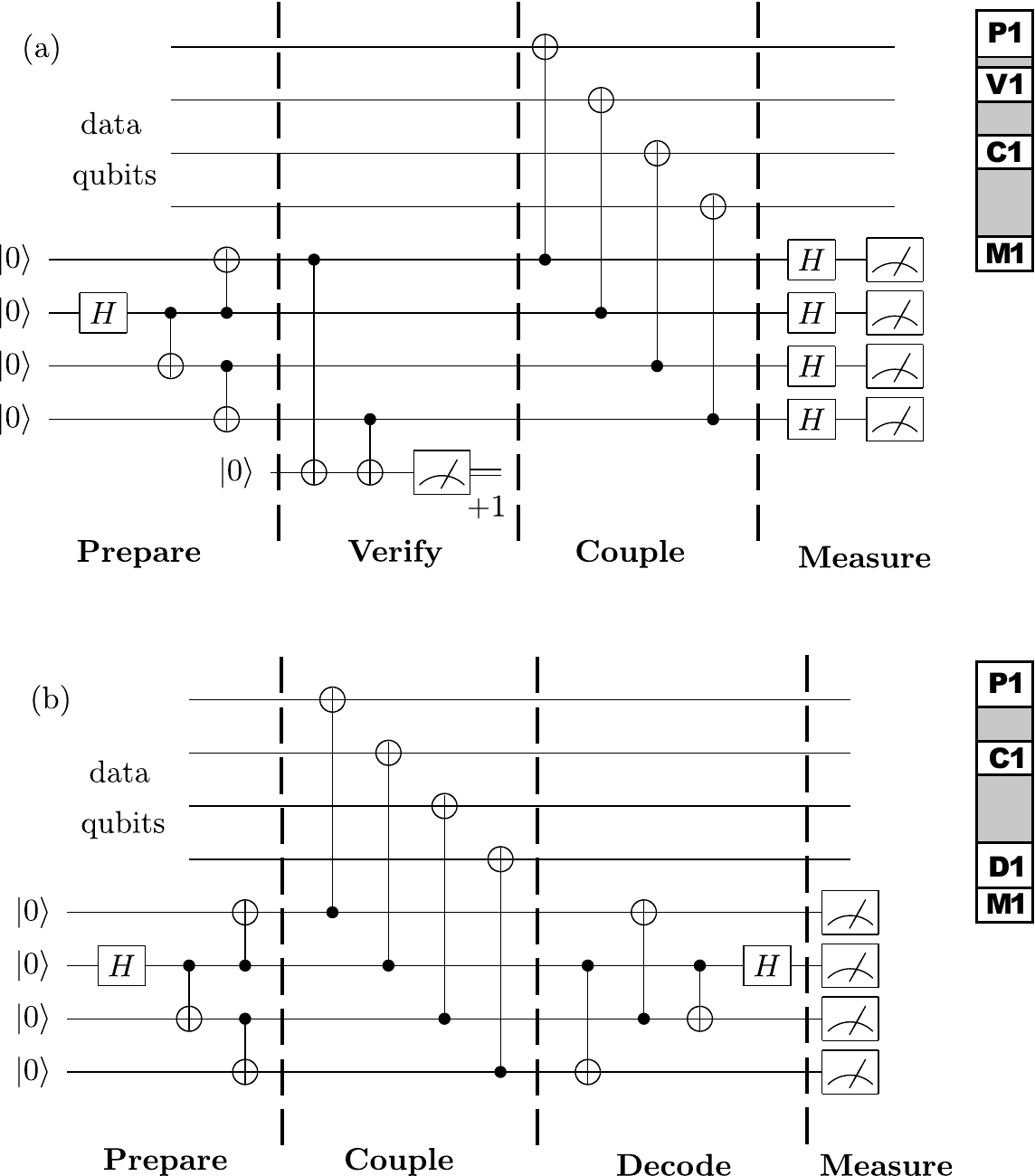}
\caption{
\label{fig:circuits} Circuits for extraction of Z type syndrome
measurement of the Steane code using the (a) standard Shor and (b) DiVincenzo-Aliferis method.  
In the DiVincenzo-Aliferis method, the cat state verification step is 
substituted with post-measurement decoding of the ancilla.
Dashed lines demarcate different sections of the circuits.
Also shown to the right are representations of the schedule of operations as a function
of circuit sections.  P=Prepare, V=Verify, C=Couple, D=Decode, M=Measure.
Grey regions correspond to operations that are exclusively movement.}
\end{figure}

The Steane code has six weight-four syndrome operators. Each syndrome
is extracted by measuring a four-qubit cat state after interacting with the data qubits. 
In this study, the Steane QEC process is simulated and the latency and 
fidelity are calculated varying numbers of Shor ancilla sets from one to six.
The two preparation/decoding cases are: (1) ``on-demand'' where only two sets of 
ancilla are prepared at any time, and (2) ``one-time,'' where all ancilla are
prepared at once before the first use.  These procedures are done for both Shor 
and DiVincenzo-Aliferis ancilla encodings. 
Circuit diagrams for the two methods are shown in Figure \ref{fig:circuits}.

\subsection{Ion trap physical machine description}

\begin{figure} \includegraphics[scale=0.5]{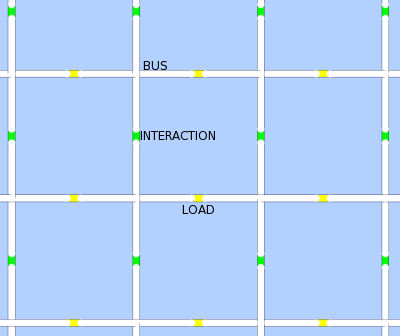}
\caption{\label{fig:topology}Layout for QCS Ion Trap PMD} \end{figure}

Previous ion trap studies in the literature have used a gate-level error model
to calculate error correction properties.  Here we model our ion trap using 
parameters and constraints derived from the Physical
Machine Description (PMD) provided by the IARPA Quantum Computer Science program~\cite{IARPA}.
The ion trap PMD is a collection of linear ion trapping regions joined by cross
junctions see Figure \ref{fig:topology}. It is modeled after the ion trap
charge-coupled device architecture of Kielpinski, Monroe, and Wineland
\cite{KMW}.  Each bus segment (white) section is capable 
of holding four ion trapping regions or ``wells.'' Each well
is capable of hosting up to five ions.  Individual ion loading wells are indicated in
yellow, and interaction wells capable of executing gate or measurement
operations are in green.  In order to undergo a two qubit gate (such as controlled-phase),
the two qubits must be co-located in an interaction well.  
Shuttling a qubit between adjacent empty wells takes 10 $\mu$s.  There is an additional time
cost of 10 $\mu$s to add or remove a qubit from a well that is occupied
This reflects the increased experimental complexity of joining and splitting single ions from ion chains \cite{RoweQIC2002,HomeQIC2006}.  

Logical errors are assumed to arise from stochastic white noise and $1/f$ noise in
the control and background Hamiltonians. The result is a very asymmetric error
model that better reflects the dominance of gate errors over memory errors in
the actual physical system. The model does not consider the heating of the ion
motion due to transport. The error of two-qubit ion gates is
modeled as a stochastic noise term in the two-qubit Hamiltonian. 
Table~\ref{tab:gateRates} gives the latency and error rate costs of each gate type for the
ion trap PMD. 

In order to get the error rates in Table~\ref{tab:gateRates}, we approximate the real error channel derived from the stochastic noise with the closest Pauli error channel.  We denote the process matrix of the noisy gate by $\chi'$.  The process matrix of the operation corresponding to the target (error-free) unitary followed by an $X$ gate is denoted by $\chi^{X}$.  
The process matrices corresponding to the target unitary followed by a $Y$ or $Z$ gate are denoted analogously.  
We then calculate the error rates as the overlap between $\chi^i$ and $\chi'$:  $X_{\text{er}} = \frac{1}{4} \langle \chi^{X}, \chi' \rangle$, $Y_{\text{er}} = \frac{1}{4} \langle \chi^{Y}, \chi' \rangle$, and $Z_{\text{er}} = \frac{1}{4} \langle \chi^{Z}, \chi' \rangle$ where  $\langle A, B \rangle = \text{Tr}(A^{\dag}B)$.    

\begin{figure} \includegraphics[scale=0.4]{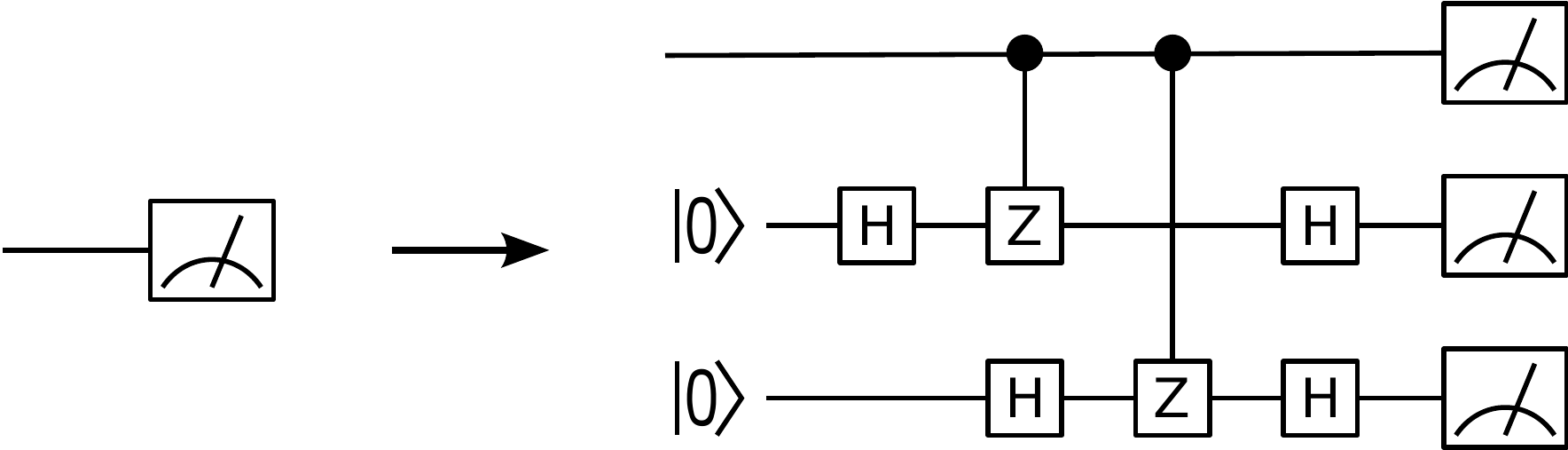}
\caption{\label{fig:multimeasure} Improvement of measurement gates by adding two
ancilla. This reduces the failure rate when the ancilla preparation and the 
controlled gates are relatively reliable compare to the measurement gates. The final
measurement value is determined by the majority vote of the three measurements. } \end{figure}

The measurement error is above the Steane code threshold but this is be fixed by introducing two
extra qubits and CZ and Hadamard gates as shown in Figure~\ref{fig:multimeasure}. 
This enhancement provides us the error rate of $O(\epsilon^2)$ where  
$\epsilon$ is the error rate of a single measurement.  The enhanced measurement operation
is denoted as `MULTIMEASURE' in Table~\ref{tab:gateRates}.

\begin{table} \caption{\label{tab:gateRates}Execution time and 
error rates of physical operations. MULTIMEASURE gates are the 
enhanced measurement gate described in Figure \ref{fig:multimeasure}.}
\begin{tabular}{l|c|c|c|c} \hline \hline Gate &   Latency (in $\mu$s) &   Error
Rate X &  Error Rate Y &  Error Rate Z    \\  
\hline 
\hline X    & 3 &  1.6E-8 &  8.0E-10   &  1.0E-9 \\
\hline Y    & 3 & 8.0E-10  &    1.6E-8 & 1.0E-9  \\ 
\hline Z    & 3 &  0.0 & 0.0  &   1.8E-8  \\
\hline S    & 2 &  0.0 &  0.0 &  5.5E-9  \\ 
\hline T    &  1 &  0.0 &  0.0 &  1.7E-9 \\
\hline HADAMARD    & 6 &  1.6E-8 & 4.0E-9 &   1.9E-9   \\ 
\hline CZ    & 105.5 &  0.0 & 0.0 &  \begin{tabular}{@{}c@{}} IZ: 6.7E-8\\ ZI: 6.7E-8\\ ZZ: 2.5E-5\end{tabular} \\ 
\hline PREPARE Z & 10 &  0.0 & 0.0 &    0.0  \\ 
\hline MEASURE Z & 100 &  0.0 & 0.0 & 1.0E-4  \\
\hline MULTIMEASURE Z & 355 & 0.0 & 0.0  & 3.1E-6 \\ 
\hline WAIT/MOVE    & $t$ &  $0.0$ & 0.0 &     $t \cdot$ 5.5E-10  \\ 
\hline JOIN/SPLIT & 10 & 0.0 & 0.0 & 5.5E-9\\
\hline \end{tabular}
\end{table}

\subsection{Quantum Machine Parameterizer} A design tool is required to model the
ion trap layout and execute qubit schedules on it.  We use the Quantum Machine 
Parameterizer (QMP) code suite developed at GTRI.  QMP is used 
for designing architecture layouts and creating operation schedules with real locality
constraints. QMP can currently be used for any hardware where the locality constraints can be mapped to a planar graph.
 QMP has three primary facets: quantum computer layout modeling,
operation scheduling, and physical qubit state tracking.

The layout modeling module allows the user to describe a quantum computing
system, specifically the physical connectivity of allowed qubit paths, and the
location of addressable zones.  
Also, this module accepts device-dependent parameters such as
gate times and movement cost times to customize the behavior of a physical
machine.  This includes such device-specific operations as 
JOIN and SPLIT operations, required for two-qubit interactions. 

The operation scheduling module allows the user to write a schedule of
operations that can be performed in a circuit-model-based quantum computer (gate
operations, qubit preparation, etc.).  This schedule is written at a
``high-level'' which we define as a list of gate operations and move requests
that only specify qubit and destination address.  The scheduling module then
calculates qubit paths using a specialized A* path-finding
algorithm~\cite{astar1971, astar1984}, parallelizes the schedule where
possible, removes possible collision events, and produces a series of qubit
movement operations. Movement parallelization is performed with highest
priority given in terms of move request order in a parallel block in the
schedule.  The first qubit is moved with no impediment, provided that a path
exists.  The second qubit's movement must defer to the first qubit, and WAIT
commands are issued as needed to the second qubit to prevent collisions.  This
continues until the end of a parallel block is reached.  In order to optimize
this approach to parallelized movement, QMP analyzes move calls, qubit start
positions, and destinations, and then re-orders the move calls as necessary.
This module uses the device-dependent timing parameters such as gate times to
complete the latency calculation of the operations schedule.  This approach
means that QMP will automatically create different (ideally optimal) 
qubit transport schedules and overall latencies for different 
choices in initial qubit arrangement, ancilla population, etc.  

Finally, the qubit state tracking module allows the user to visualize the
positions of the physical qubits within the layout as a function of time, and
produce as output the total latency of the operations schedule.  The module
also produces an ``error schedule'' file which reduces the detailed physical machine
schedule to a sequence of error-relevant events that the Quantum Circuit Fault
Tracer (Section \ref{subsec:QCFT}) uses to calculate failure rate.

\subsection{\label{subsec:QCFT}Quantum Circuit Fault Tracer}

The Quantum Circuit Fault Tracer (QCFT) is a tool created to efficiently compute logical
failure rates of concatenated FTQEC codes. This tool is based on the 
``fault paths'' concept, introduced by Aliferis, Gottesman, and Preskill 
to calculate thresholds of distance-3 concatenated FTQEC codes~\cite{Aliferis2006}. 
It takes as input a quantum circuit and failure
rates of physical gates including WAIT and MOVE. Starting from the output state
qubits, the circuit is traced backwards, marking possible fault points. Fault points
are circuit locations where an error can propagate into a fatal error
for the FTQEC circuit. The QCFT then combines the fault points and calculates
the overall failure rate of the given circuit. To improve the accuracy of the
output logical failure rate, we separate errors into X-type and Z-type, and
propagate them on the circuit independently. Each Y-type error rate in 
Table~\ref{tab:gateRates} is added into both X-type and Z-type rates. 
When the input circuit is a distance-3 QECC circuit, the output is the 
probability of having errors propagated into two or more output data qubits.


\section{\label{sec:Proc}Procedure}

\begin{figure}[hbtp]
\center
\includegraphics[scale=0.7]{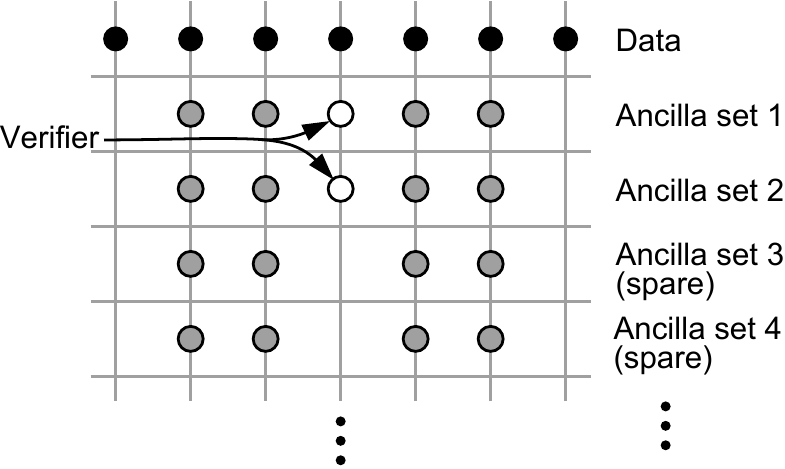}
\caption{Sample layout for the Steane algorithm with four sets of Shor ancilla and
two preparation rows.  The data qubits never move from their positions in the top row.
The preparations rows are indicated by the presence of 
static verifier qubits.  \label{fig:layout}} \end{figure}

Executing an algorithm on the ion trap layout requires a set of
starting positions for all of the qubits present in the computation. 
Each starting position corresponds to an interaction well.  That is,
each qubit has its own home interaction well which it starts in at the
beginning of the QEC round; see Figure~\ref{fig:layout}.
The qubits are ordered into rows by function.  The data qubits sit in the top row, and
never move from their initial positions.  Each additional row of qubits is a self-contained
ancilla set, which is prepared according the Steane-Shor or Steane-DiVincenzo-Aliferis circuit,
coupled with the data, returned to the ancilla area
(although not necessarily to the set's original position) and measured to obtain
the error syndrome.  
  Each qubit returns to its home well after being
called away for a sequence of two-qubit gates.  By this convention, the control
bit travels and the target bit stays home.  These choices remove some optimization
capability, but allow us to test and compare different QEC circuits from
the same initial conditions and scheduling assumptions.

Once the initial state and scheduling assumptions are established, 
QMP is used to calculate the time required to perform level 
one error correction assuming different operation times and ancilla 
management strategies. This includes varying the method of ancilla 
preparation and measurement, the number of ancilla, the parallelization 
of ancilla manipulation, and the time of gates and measurements.  For 
each set of conditions, QMP uses the A* algorithm to optimize the 
latency from an initial hand-crafted schedule.  An error schedule is then
produced containing all gate and latency information.
Finally, QFCT reads the error schedule 
and determines the logical error rate for a set of these conditions. 

The Steane code is theoretically improved in terms of 
latency and error rate by creating multiple ancilla sets in parallel. 
To compare the efficacy of ``on-demand'' ancilla with ``one-time'' ancilla,
we look at two different parallelizations for both Shor and DiVincenzo-Aliferis ancilla.
At this point, we use the following 
notation to represent choices of ancilla management: $y$P$x$R 
where $y$ is the number of ancilla sets that can be prepared and measured 
simultaneously ($y$=All means complete parallelization over the set) 
and $x$ are the number of ancilla sets, one set per row in the layout.

The on-demand approach is 2P$x$R where ancilla preparation is only allowed in the two rows
immediately below the data row. In this arrangement, ancilla qubits are moved up
into one of the preparation rows, prepared, coupled with the data, and then
moved to the bottom of the ancilla ``stack'' for measurement, which makes room
for the next ancilla set.  For the Shor case, 
verifier qubits are only kept in the two preparation rows.  Six sets of ancilla
are prepared in total in order to perform the three bit-stabilizer and three 
phase-stabilizer measurements.

The one-time approach is AllP$x$R wherein all ancilla are prepared
at once and coupled with the data as soon as possible. For fewer than six ancilla sets,
the ancilla sets are prepared at once, coupled with the data, measured, and then
prepared again as soon as possible, repeated until all stabilizer measurements
are performed.  For the Shor case, every row has a 
verification qubit.  

\section{\label{sec:Res}Results and Discussion}
\subsection{Scheduled error correction: Two-row preparation}

\begin{figure}[hbtp] \center
\includegraphics[scale=1]{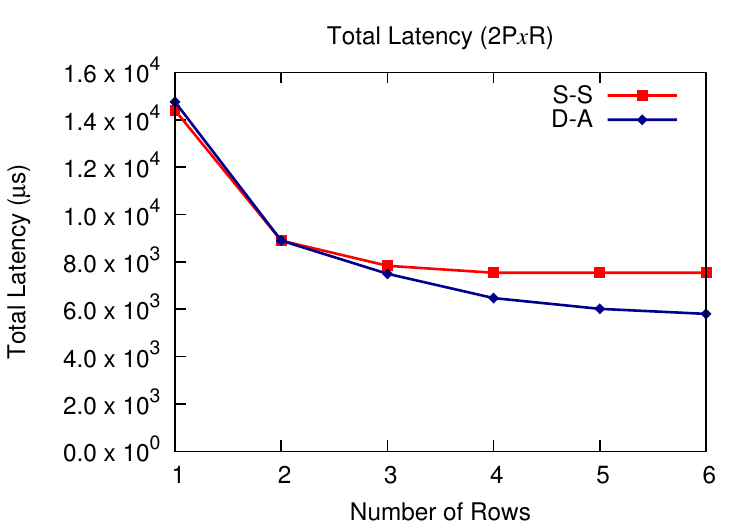} \caption{Total latency
for the Steane-Shor(S-S) and Steane-DiVincenzo-Aliferis(D-A) algorithms
as a function of number of ancilla sets for the case of two preparation rows
(2P$x$R). \label{fig:defTimes2}} \end{figure}

The total latencies for the Steane-Shor and Steane-DiVincenzo-Aliferis algorithms
are shown in Figure~\ref{fig:defTimes2}.  The
latency for one ancilla set represents the standard one-set strategy for the
algorithms.  For the case of a single ancilla row, the latencies for 
Steane-Shor and Steane-DiVincenzo-Aliferis are almost identical, with a slight
advantage to Steane-Shor. In terms of latency, the verification step in Steane-Shor is roughly
equivalent to the decoding step in Steane-DiVincenzo-Aliferis.  However, for Steane-Shor,
ancilla qubits can be moved to the data qubits while the verifier qubit is being 
measured.  By contrast, the ancilla qubits are tied up during decoding, and no
further parallelization is possible.  Thus, assuming the verifier show successful
ancilla creation, the Shor encoding is slightly more efficient.  
As will be shown, increasing the gate time, increasing the 
measurement time, or adding additional ancilla rows (increasing parallelizability) 
will separate Steane-Shor and Steane-DiVincenzo-Aliferis performance.

The latencies decrease greatly for both encodings with the addition of a second
preparation row (2P$x$R) and, in general, continue to decrease with the addition of
spare ancilla sets.  The Steane-Shor latency reaches a
minimum with four ancilla sets (two spare sets).  This occurs because the
preparation step in the Steane-Shor algorithm dominates the time
required to perform a single bit or phase stabilizer measurement.  
Since in this scheme, only two sets can be 
prepared at a time, four total ancilla sets is sufficient to continuously utilize the preparation rows.

The latency for the Steane-DiVincenzo-Aliferis algorithm reaches a minimum with a full
six sets of ancilla.  This occurs because, unlike preparation, decoding can occur
at any interaction well.  Additional ancilla rows allow for greater parallelized preparation
and decoding and consequently less down-time between data-coupling steps.
The elimination of the limited verifier measurement
in mid-circuit clears out a critical bottleneck in the QEC execution.

\begin{figure}[hbtp] \center
\includegraphics[scale=0.7]{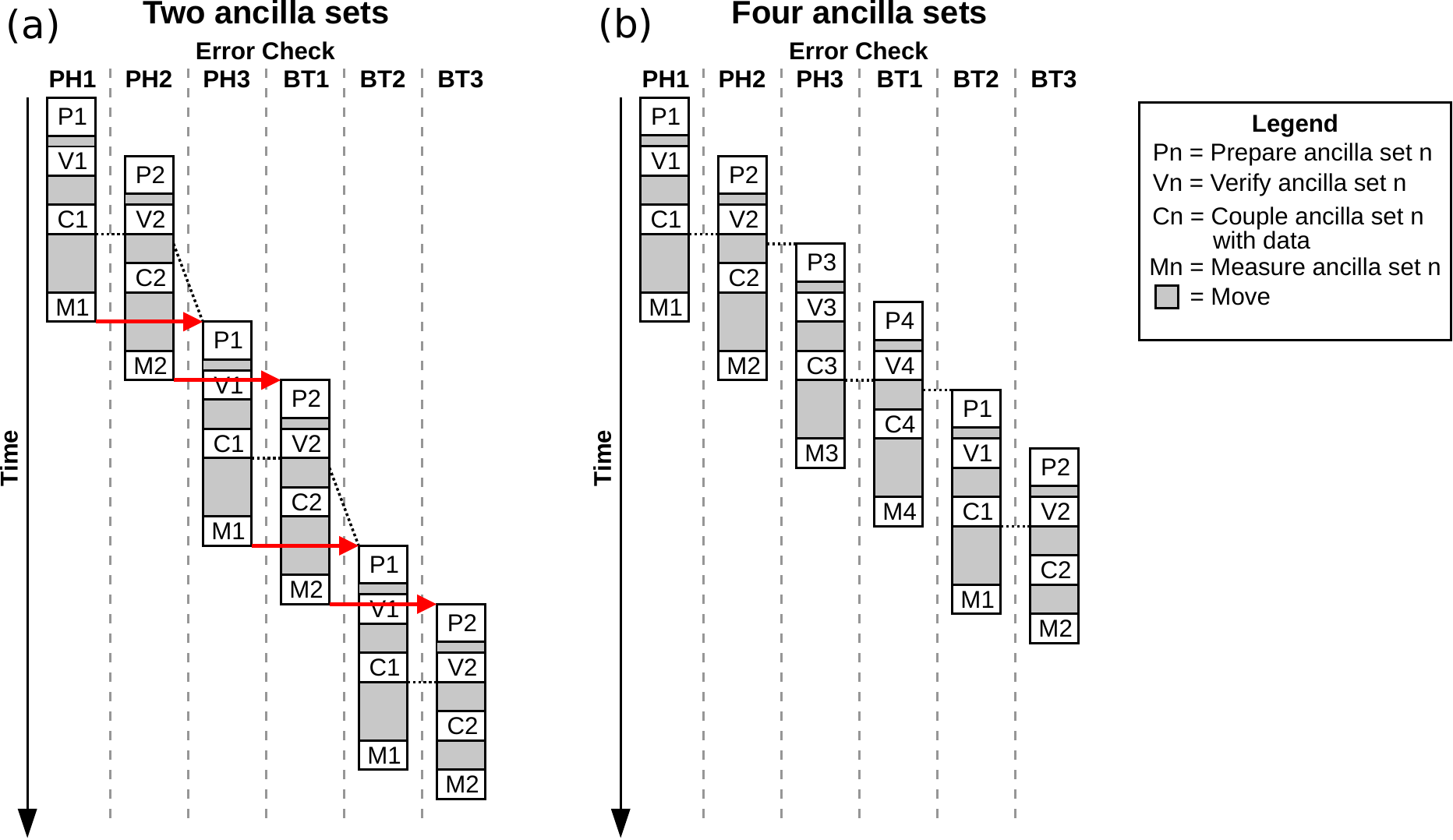} \caption{Parallel
strategy for implementing the Steane-Shor algorithm with (a) two sets of ancilla
and two preparation rows and (b) four sets of ancilla and two preparation rows.
  ``PHx'' and ``BTx'' indicate bit-stabilizer or phase-stabilizer
operations, respectively.  Dashed lines indicate steps for separate ancilla
sets that must occur in sequence. Red arrows
indicate steps that must occur in sequence because an ancilla set is reused.
The need to reuse the two ancilla sets, as indicated by the red arrows in (a), 
prevents the first set from being prepared as early as possible, as indicated by the
non-horizontal dashed lines between PH2 and PH3 and BT1 and BT2.  The additional
ancilla sets in (b) ensure that the ancilla are used at the speed of computation,
with measurement of an ancilla set occurring before the need to prepare that
ancilla set.
\label{fig:tc2}} \end{figure}

More specifically, the efficiency of the algorithm execution depends 
on the degree of overlap between separate operations.
The individual stabilizer measurements must be performed sequentially when using the
one-set strategy.  With two rows of ancilla, two stabilizer measurements can be performed
in parallel.  This is achieved by overlapping partially the preparation of the
second set with the first.  The preparation of the second set is delayed so
that moving it to the data takes place when the first set is moved back to the
top row for measurement.  Following coupling with the data, the second set is
moved to its original row via the unoccupied outer columns of the layout for measurement.  Once
a set of ancilla is measured, it is re-prepared for its next stabilizer measurement.
This process, shown in Figure~\ref{fig:tc2}(a), is repeated two more times to
perform the remaining four stabilizer measurements.

Adding a third or fourth set of ancilla provides additional spare sets that can move up and
begin preparation as soon as the second set moves to the data.  Furthermore,
the measurements of the first and second sets complete before it is necessary
to reuse them to perform the final two stabilizer measurements.  Therefore, 
unnecessary delays are removed by using more sets.  This process is
shown in Figure~\ref{fig:tc2}(b).  For Steane-Shor, adding fifth and sixth sets removes the need to
reuse ancilla but does not provide any further decrease in the latency, due
to the verification bottleneck.

\begin{figure}[hbtp] \center
\includegraphics[scale=0.7]{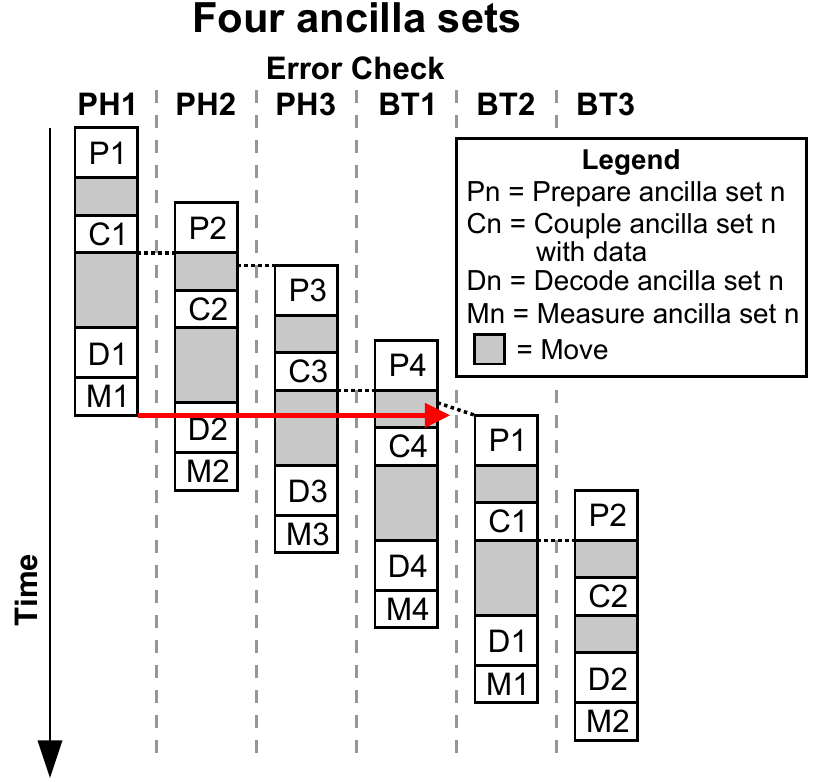} \caption{Parallel
strategy for implementing the Steane-DiVincenzo-Aliferis algorithm with four sets of
ancilla and two preparation rows. The need to reuse ancilla set one, as
indicated by the red arrow, does not permit this set to be prepared for BT2 as
early as possible, as indicated by the non-horizontal dashed line between BT1
and BT2. \label{fig:tc4d}} \end{figure}

Comparing DiVincenzo-Aliferis to Shor shows that Steane-DiVincenzo-Aliferis
requires less time for preparation but more time for measurement.
As a consequence, the ancillae cannot be reused as quickly
as they are needed for a subsequent preparation.  For example, with four sets
of ancillae, the first four stabilizer measurements can be performed in rapid succession.
However, ancilla sets three and four are prepared quickly and move to the data
before the decoding/measuring of sets one and two are complete.  Thus, the
preparation of sets one and two for the last two stabilizer measurements cannot begin as
early as possible.  This process is shown in Figure~\ref{fig:tc4d}.  A unique set
of ancilla must be available for each stabilizer measurement operation (six sets) to
optimize fully the parallelization of the DiVincenzo-Aliferis algorithm.

\subsection{Scheduled error correction: All-row preparation}

Further decreases in the latency for the Steane-Shor and Steane-DiVincenzo-Aliferis
algorithms can be achieved when three or more sets of ancillae are used by
preparing every set in parallel.  This all-row preparation allows as many 
stabilizer measurements as there are ancilla sets to be performed in rapid succession.  However,
preparing on three or more rows introduces delays between preparing the lower
sets of ancillae and coupling them with the data.  
Here, the performance of the all-row strategy is
studied for two, three, and six sets of ancillae.  Four and five sets of ancilla are
not considered because they are not commensurate with the total number of 
stabilizer measurements that must be performed.

\begin{figure}[hbtp] \center
\includegraphics[scale=1]{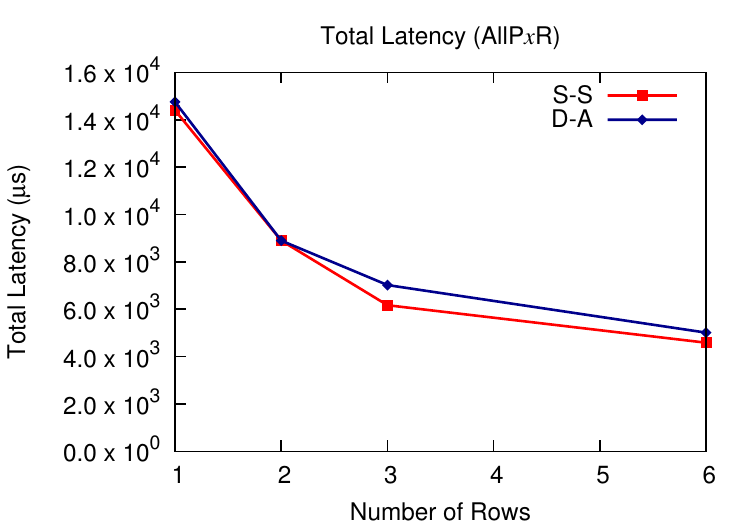} \caption{Total
latency for the Steane-Shor(S-S) and Steane-DiVincenzo-Aliferis(D-A) 
algorithms as a function of number of ancilla sets. \label{fig:defTimesA}}
\end{figure}

The total latencies for the Steane-Shor and Steane-DiVincenzo-Aliferis algorithms 
for all-row preparation are shown in Figure~\ref{fig:defTimesA}.
The latencies again decrease
consistently with the addition of extra ancilla sets, due to the 
ability to prepare and measure ancilla sets in parallel.  The biggest change 
from ``on-demand'' ancilla is that for Shor method, all ancilla rows are allowed
to have verification.  This removes the bottleneck seen in the two-row
case.  The parallelization advantage of Steane-Shor (moving ancilla qubits
to data qubits for coupling while verification measurement is occurring)
then gives it a lower latency than DiVincenzo-Aliferis for any number of 
ancilla sets.

\subsection{Gate time variation}

\begin{figure}[hbtp] \center
\includegraphics[scale=1]{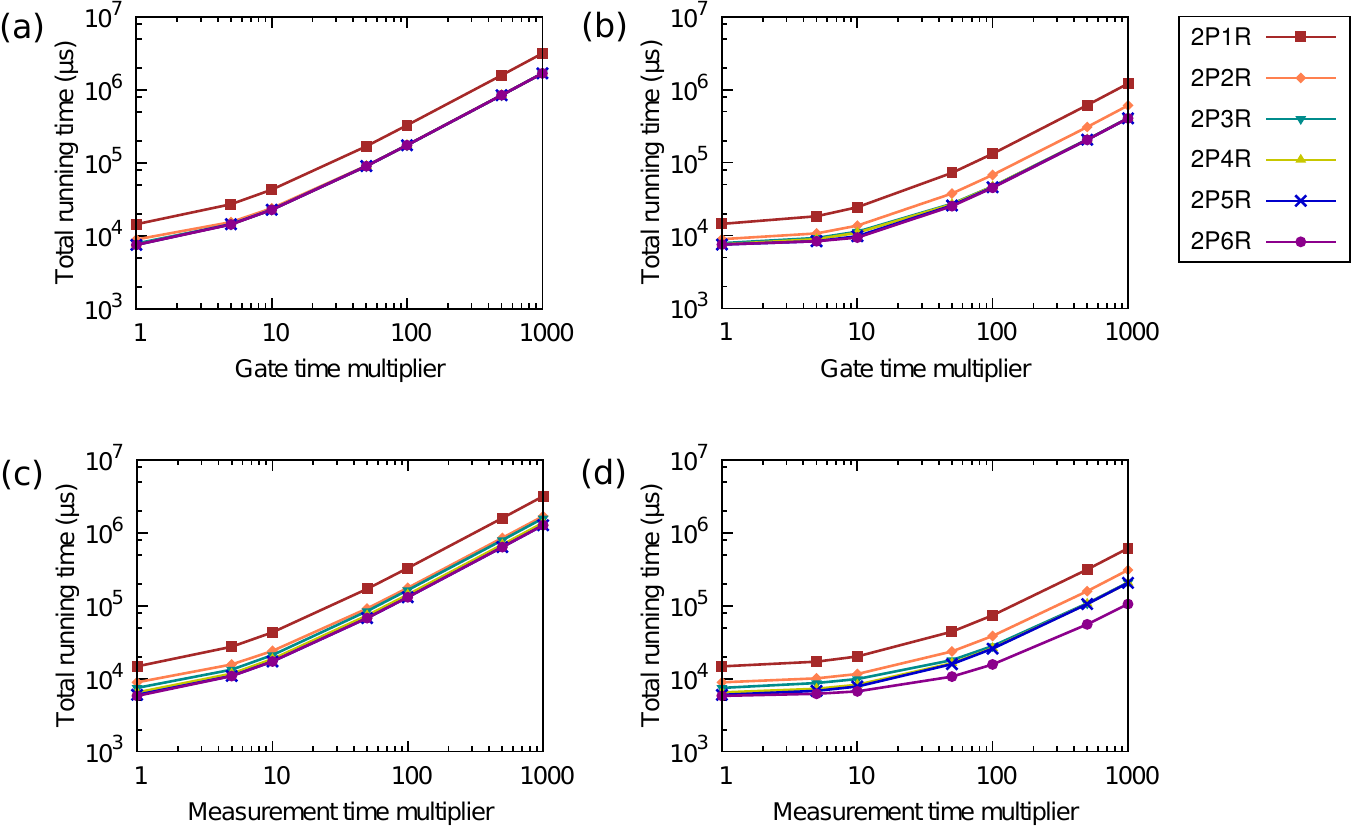} \caption{
Top: Total latencies for the (a) Steane-Shor and (b) Steane-DiVincenzo-Aliferis algorithms as a
function of gate-time multiplier for various numbers of ancilla sets.
Bottom: Total latencies for the (c) Steane-Shor and (d) Steane-DiVincenzo-Aliferis algorithms as a
function of measurement-time multiplier for various numbers of ancilla sets.  In all cases,
preparation is limited to the top two ancilla rows.
\label{fig:gateTimes2}} \end{figure}

The total latencies and the effectiveness of using spare ancilla sets to
parallelize the algorithms vary with the gate times.  For the case of two-row preparation, the latencies of the Steane-Shor and Steane-DiVincenzo-Aliferis algorithms as a function of a gate-time
multiplier for various numbers of ancilla sets are shown in
Figure~\ref{fig:gateTimes2}(a) and (b) respectively.  Latencies
asymptotically approach a linear dependence on the gate time for large gate
times.  This is expected because the time spent on moves and measurements
becomes negligible compared to the time spent on gates specifically controlled
gates.

As the gate times get larger and dominate the
total latency, run time improvement is dependent on the time spent on CNOT
gates.  The one-set strategy for the Steane-Shor algorithm requires 30 CNOT
stages to be performed, where a stage is defined as one or more overlapping
CNOT gates.  In contrast, the two-row preparation strategy requires only 16 CNOT
stages, dropping the latency almost in half.  This is because more CNOT gates 
can be performed in overlapping pairs
or triplets.  The two-row preparation limit prevents any further significant
latency reduction, since additional rows have to wait for a preparation row to clear out
before they are prepared.  In particular, having three ancilla sets reduces Steane-Shor to
14 CNOT stages, a very minor improvement over two ancilla sets, while having four to six
ancilla sets offers no additional reduction beyond three sets.  

\begin{figure}[hbtp] \center
\includegraphics[scale=1]{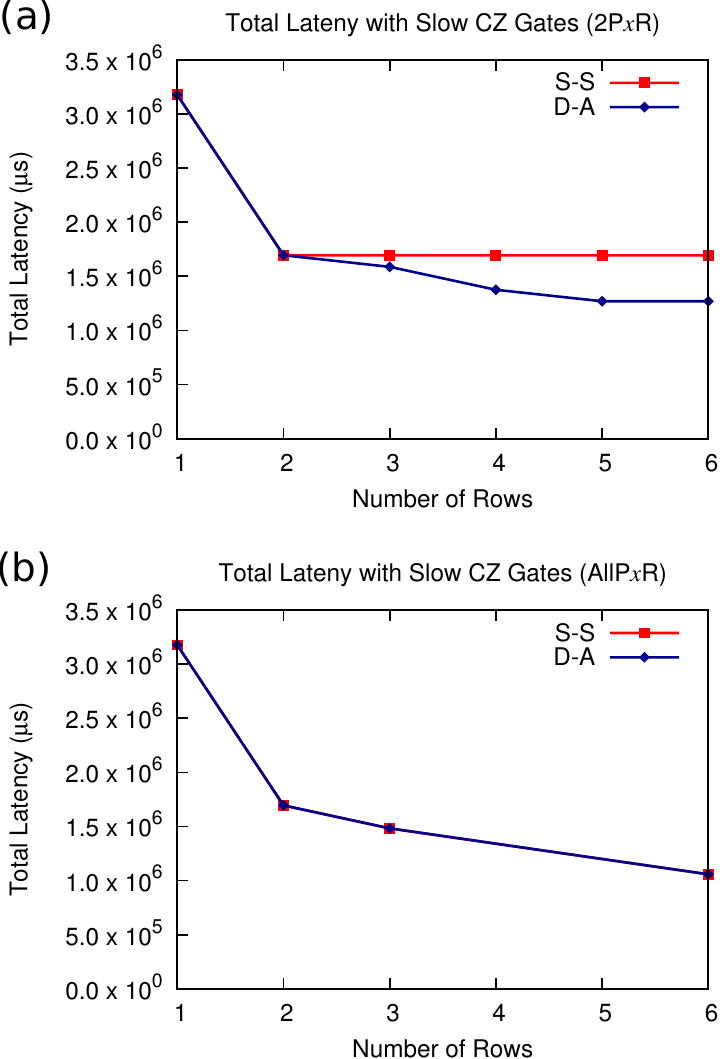}
\caption{
Latency for Steane-Shor(S-S) and Steane DiVincenzo-Aliferis(D-A) circuits
as a function of total number of ancilla rows for the case
of CNOT execution time 1000 times longer than the default time.  
(a) Two preparation rows (2P$x$R); (b) ancilla can be prepared on all rows (AllP$x$R).
}
\label{fig:longGate}
\end{figure}

For the DiVincenzo-Aliferis algorithm in the case two preparation rows, 
time saved with two ancilla sets follows a similar trend to that achieved with
Steane-Shor algorithm.  However, using more sets
provides greater latency reduction since decoding is not subject
to the two-row preparation limit.
Similar to the
Steane-Shor algorithm, the one-set strategy for the Steane-DiVincenzo-Aliferis algorithm
requires 30 CNOT stages.  Two rows reduces this to 16 CNOT stages, three rows to
12 stages, four rows to 11 stages, and five or six rows to 10 stages.  Six rows
reduces the latency a slight additional amount due to better parallel transport.
The diminishing returns in adding ancilla rows for the case of long gate times 
can be seen in Figure~\ref{fig:longGate}(a).

\begin{figure}[hbtp] \center
\includegraphics[scale=1]{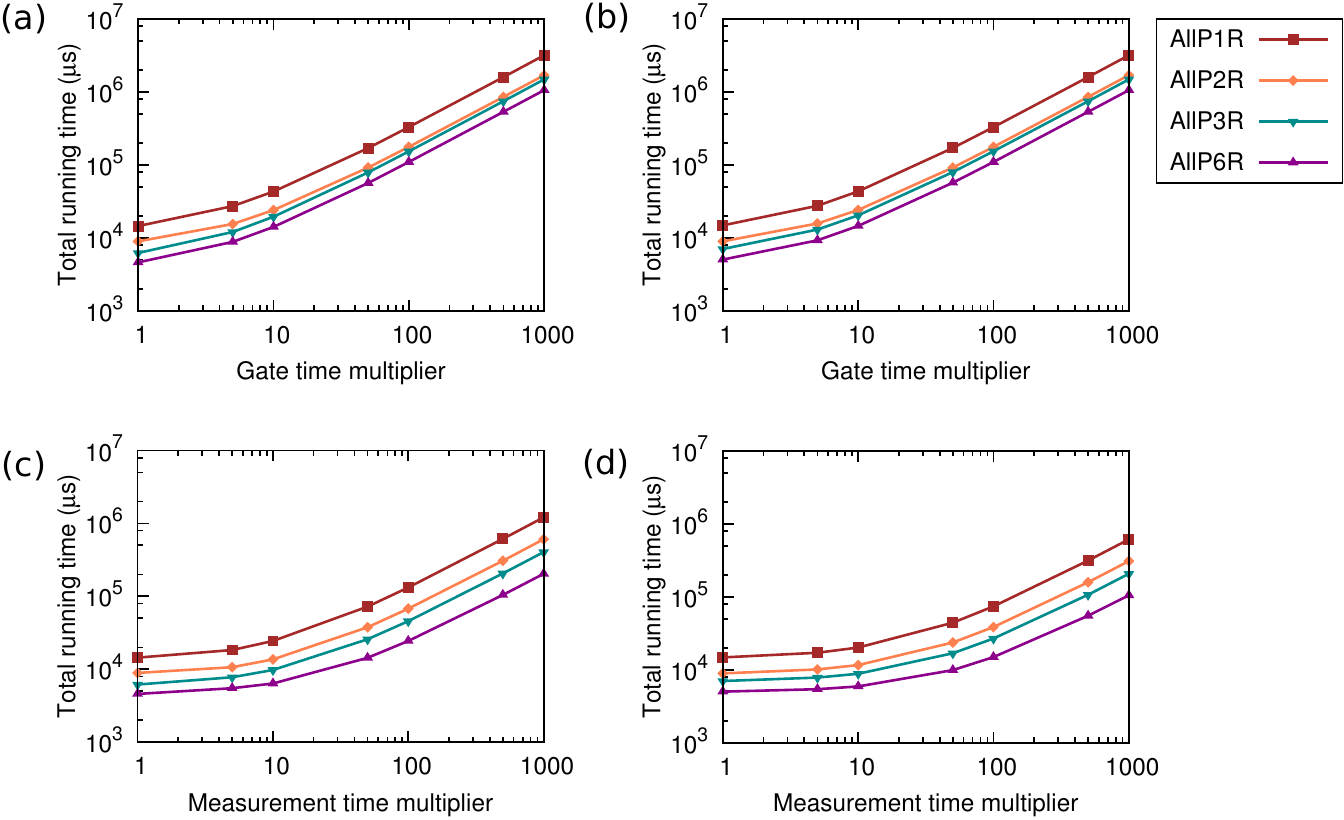} \caption{
Top: Total latency for (a) the Steane-Shor and (b) Steane-DiVincenzo-Aliferis algorithms as a
function of gate-time multiplier for various numbers of ancilla sets. 
Bottom: Total latency for (c) the Steane-Shor and (d) Steane-DiVincenzo-Aliferis algorithms as a
function of measurement-time multiplier for various numbers of ancilla sets. In all cases,
the ancilla sets are prepared at-once (no limit on preparation).
\label{fig:gateTimesA}} \end{figure}

All-row preparation results as a function of a gate-time
multiplier for various numbers of ancilla sets are shown in
Figure~\ref{fig:gateTimesA}(a) and (b) respectively.  These times again
asymptotically approach a linear dependence on the gate time for large gate
times, because of the dominance of gate time in the latency.  In this case,
Steane-Shor and Steane-DiVincenzo-Aliferis are nearly identical in latency.  Without
the preparation limit, both approaches have the exact same number of CNOT stages
for any given number of ancilla sets.  Just as in the case of all-row preparation
for default gate times described above, Steane-Shor has slightly lower latency
due to parallelizing transport and verifier measurement.  As the gate time is
increased, this latency difference becomes vanishingly small.  This behavior
can be seen in Figure~\ref{fig:longGate}(b).

\subsection{Measurement time variation}

The total latency and the effectiveness of using spare ancilla sets to
parallelize the algorithms also vary with the measurement time.  The results
for the two-row-preparation Steane-Shor and Steane-DiVincenzo-Aliferis algorithms as a function of a
measurement time multiplier for various numbers of ancilla sets are shown in
Figure~\ref{fig:gateTimes2}(c) and (d) respectively.  Results for all-row-preparation
are shown in Figure~\ref{fig:gateTimesA}(c) and (d). 
As with increased gate time,
the latency is dominated by measurement time for large
measurement times.  This is expected because the time spent on moves and gates
becomes negligible compared to the time spent on measurements.

For both preparation types, the latency is reduced
consistently as more sets of ancilla are used for each algorithm.  As with the
gate time variation, this is because more stabilizer measurements can be performed in
parallel with more sets.  The time saved also increases consistently as the
measurement time increases.  This increase indicates that more measurements are
performed in parallel.  

As in all other cases, Steane-Shor sees a saturation effect as the number
of rows are increased.  Increasing the measurement time puts more emphasis
on the verification measurement bottleneck, since the ancilla-data controlled gate
is not allowed to occur until after the verification measurement is completed.
For extremely long measurement times and one ancilla set, Steane-Shor is effectively
12 measurement stages- one verification measurement and one ancilla measurement
per stabilizer.
For two ancilla sets, Steane-Shor is effectively 
7 measurement stages.  For three or more ancilla sets, it saturates
at 4 measurement stages.  For shorter measurement times, this behavior is moderated
by transport and gate times, but the latency reduction still appears.

\begin{figure}[hbtp] \center
\includegraphics[scale=1]{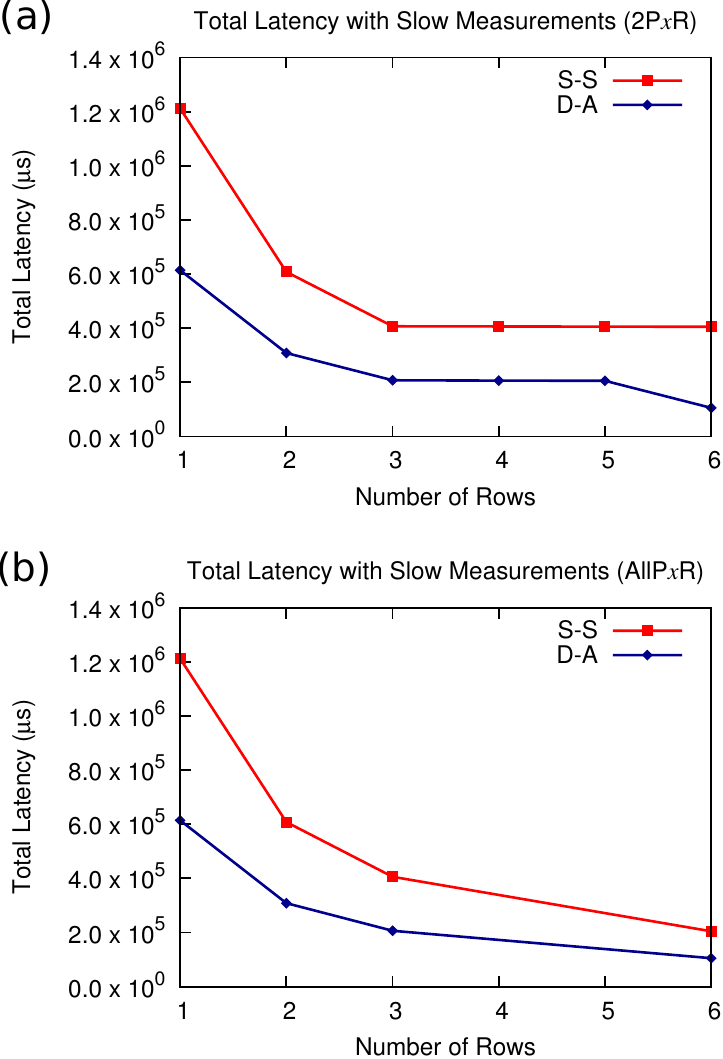}
\caption{
Latency for Steane-Shor(S-S) and Steane-DiVincenzo-Aliferis(D-A) circuits
as a function of total number of ancilla rows for the case
of measurement execution time 1000 times longer than the default time.  
(a) Two preparation rows (2P$x$R); (b) ancilla can be prepared on all rows (AllP$x$R).
}
\label{fig:longMeas}
\end{figure}

By contrast, for long measurement times, Steane-DiVincenzo-Aliferis sees a 
superior latency for even a single ancilla row, and a staggered decrease in latency for
additional ancilla sets.  This is not surprising, since handling long measurement
times was the motivation for this scheme.  For a single set and extremely long
measurement times, Steane-DiVincenzo-Aliferis is
effectively 6 measurement stages (one measurement per stabilizer).  
Using the two-row preparation strategy with two ancilla sets, measurement
is reduced to 3 stages.  Using three, four, or five sets reduces this to 2 measurement 
stages, and finally six sets reduces the latency to 1 measurement stage.  This behavior
simply comes from dividing the total number of measurement stages (six)
by the number of ancilla sets, rounding up. Figure~\ref{fig:longMeas}(a) shows
the behavior of both encodings for long measurement times (1000 times longer than
the default value).

For the case of long measurement and all-row preparation, both Steane-Shor and 
Steane-DiVincenzo-Aliferis follow
the same latency reduction: the algorithm is reduced to twelve or six measurement
stages, respectively, which is divided by the number of number of ancilla sets
rounded up.  In all cases, DiVincenzo-Aliferis is superior to Shor method in terms of latency.
This is shown in Figure~\ref{fig:longMeas}(b).

\subsection{Scheduled error correction: Two-row versus all-row preparation}

\begin{figure} \begin{center} \begin{tabular}{ccc}
\resizebox{50mm}{!}{\includegraphics{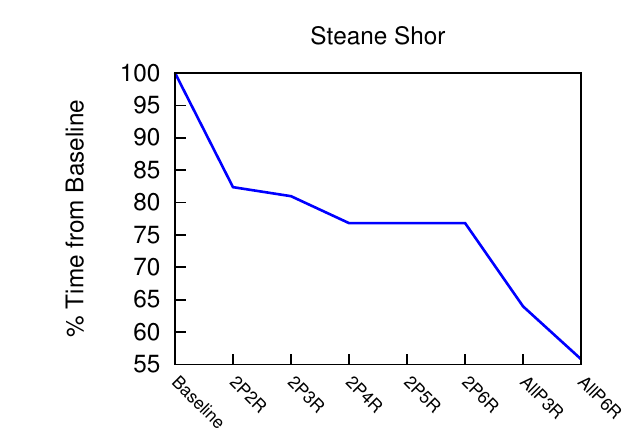}} &
\resizebox{50mm}{!}{\includegraphics{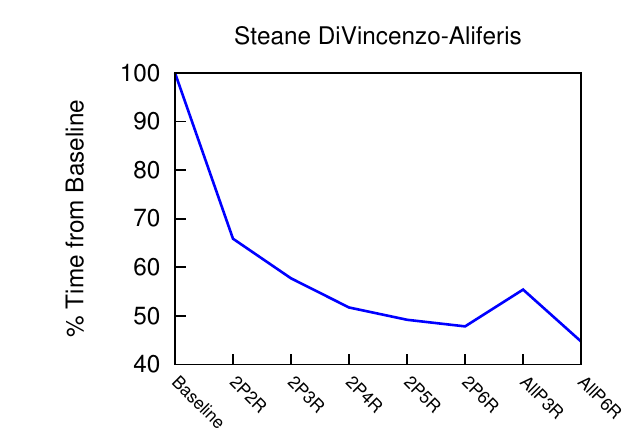}} & 
\resizebox{50mm}{!}{\includegraphics{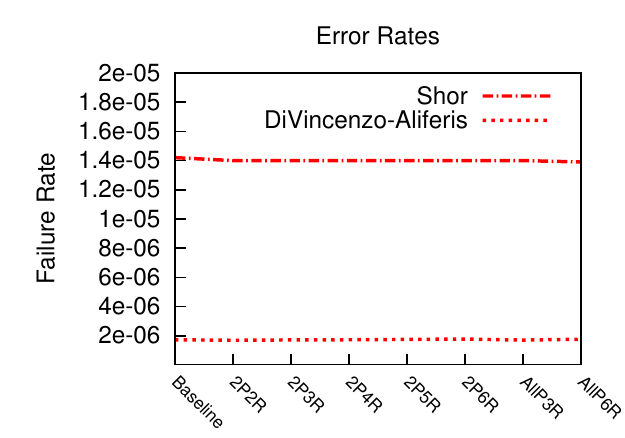}} \\

    \end{tabular} \caption{\label{fig:Res}Steane Shor and Steane-DiVincenzo execution
time and error rates. Baseline schedule keeps only one set of ancillae and
re-uses them by preparing it six times. We assume that the whole set of syndrome extraction is 
repeated three times to reduce the measurement errors. Execution times are shown 
relative to the baseline which are $5.1\times10^4 \mu s$ for the Steane-Shor circuit and 
$5.2\times10^4 \mu s$ for the Steane-DiVincenzo-Aliferis circuit.
} \label{test4} \end{center}
\end{figure}

Figure~\ref{fig:Res} shows the total execution time and logical error rates of Steane
QECC with different numbers of ancilla qubits and their scheduling scheme.
In order to to reduce the errors in syndrome measurements, we assume that we run 
the whole QECC circuits three times and the final syndromes are determined by the 
majority vote. This enables us to ignore one measurement error on a set of 
syndrome extraction.

As expected, it takes the longest to execute the whole QECC scheme when we keep
only one set of ancilla in both Steane-Shor and Steane-DiVincenzo-Aliferis circuits. For
Steane-Shor circuits, the number of preparation rows has the most influence on
the execution time. Adding four more ancilla sets with 16 qubits only reduced
the time by an additional 5$\%$ (2P2R $\rightarrow$ 2P6R), while adding more
preparation rows with the same numbers of qubits reduced the execution time an
additional 15 to 20$\%$  (2P3R $\rightarrow$ AllP3R and 2P6R $\rightarrow$
AllP6R). The execution time of the Steane-DiVincenzo-Aliferis circuits are more
susceptible to numbers of available ancilla qubits. Adding four ancilla sets
with the same number of preparation rows reduces the total time by  20$\%$
(2P2R $\rightarrow$ 2P6R).

The logical error rates are, however, almost constant regardless of number of
ancilla qubits and preparation rows. We see that the errors on the measurements 
and the two-qubit gates dominate over the error on the idle and moving qubits.
For example, the memory error rate on a qubit being idle for the total execution 
time of baseline Steane-Shor schedule would be $2.75\times10^{-5}$. This is comparable 
to the error rate of a single CNOT gate, and each qubit in a Steane QECC encounters
multiple CNOT gates. We also find that ancilla decoding (Steane-Divincenzo-Aliferis) yields a substantially lower error than ancilla verification (Steane-Shor). An abstract model using Steane syndrome extraction, instead of Shor syndrome extraction, also showed a fidelity improvement when decoding was used instead of verification \cite{ nada2013}.

Taken in total, the results suggest that increasing the number of simultaneous
ancilla preparations has the greatest impact on QEC run times, without 
adversely affecting the QEC error rate.  It also shows that the Steane-DiVincenzo-Aliferis
algorithm is equivalent or superior to the Steane-Shor algorithm, particularly for long 
measurement times, given the error model presented in this paper.  This is
attributable to reducing (for simultaneous preparation) or removing (for DiVincenzo-Aliferis) 
the ancilla verification bottleneck after preparation.  

\section{\label{sec:Con}Conclusion}

We examined changes in execution time and logical error rates of Steane QECC by
varying the number of ancilla qubits and how they are scheduled on the ion trap
architecture. We identified possible resource bottlenecks and opportunities for parallelism  
in preparing blocks of Shor ancilla.  
After studying both standard Shor and DiVincenzo-Aliferis ancilla for a variety of 
multiple ancilla set preparations, we found that one-time ancilla preparation was superior to 
on-demand preparation.  This is attributed to the time-intensive process of 
ancilla preparation driving QEC latency.  On-demand ancilla preparation limits
the speed of the QEC round to the speed of sequential ancilla preparations,
particular for verification schemes like Steane-Shor.  
In comparing the Steane-Shor and Steane-DiVincenzo-Aliferis latencies, we found that 
for the case of a single ancilla set with roughly equivalent gate, measurement,
and transport times, Steane-Shor has slightly lower latency.  This is due to
the ability to perform parallel operations on the verification and ancilla
qubit.  This also holds true for multiple ancilla sets with one-time preparation.
For on-demand preparation, verification becomes a bottleneck significantly slowing down 
Steane-Shor compared to Steane-DiVincenzo-Aliferis as ancilla sets are added.  When
gate times are increased, Steane-Shor and Steane-DiVincenzo-Aliferis become effectively
identical, as they both have the same number of parallel CNOT operations.  As 
measurement times are increased, Steane-DiVincenzo-Aliferis shows a much lower latency,
as expected.

The results presented are based on an ion trap description with optimistic error 
rates, but pessimistic gate and movement times.  The long times required for error 
correction in this paper could be improved in a number of ways.  For example by 
using ultrafast lasers,  single qubit gate times as fast as  50 ps have been 
achieved \cite{CampbellPRL2010} and two-qubit gate times can in principle  be 
considerably improved \cite{GZC2qubit,Kielpinski2qubit}.  Transport in the model 
is limited to 1 m/s but recent experiments have shown that a transport speed of 
40-80 m/s can be achieved while still controlling the quantum states of the ion 
motion \cite{BowlerPRL2012, WaltherPRL2012}. 

In the future, we plan to extend this analysis to other quantum error-correcting 
codes including non-CSS type codes. We can also apply the same methods to study 
performance of various quantum architectures and to determine whether there exists affinity
between certain devices and types of codes. The QMP method is flexible enough to
handle a wide array of qubit architectures and couplings.  
Future improvements to the QCFT 
method will allow us to approximate error rates for circuits beyond those 
generated by Clifford operators.

\begin{acknowledgments}

This work was supported by the Office of the Director of National Intelligence
- Intelligence Advanced Research Projects Activity through Department of
Interior contract D11PC20167. Disclaimer: The views and conclusions contained
herein are those of the authors and should not be interpreted as necessarily
representing the official policies or endorsements, either expressed or
implied, of IARPA, DoI/NBC, or the U.S.  Government.

\end{acknowledgments}


\begin{thebibliography}{40}
\bibitem[{IAR()}]{IARPA}
\bibinfo{note}{The ion trap PMD is part of the government furnished information
  that IARPA has provided for official use only. The necessary data to
  reproduce the results are contained in this paper.}
\end{thebibliography}


\end{document}